\documentclass[fleqn,usenatbib]{mnras}

\usepackage{newtxtext,newtxmath}

\usepackage[T1]{fontenc}

\usepackage{graphicx}	
\usepackage{amsmath}	
\usepackage{mathtools}

\usepackage{amssymb}	
\usepackage{booktabs}
\usepackage{subcaption}
\usepackage{float}
\usepackage{array} 

\include{new_commands}
\newcommand{\ML}{ML21}

\title[Long-term instability of the inner Solar System]{Long-term instability of the inner Solar System: numerical experiments}

\author[N. H. Hoang, F. Mogavero, J. Laskar]{
Nam H. Hoang\thanks{E-mail: nam.hoang-hoai@obspm.fr},
Federico Mogavero\thanks{E-mail: federico.mogavero@obspm.fr},
Jacques Laskar \\
IMCCE, CNRS UMR 8028, Observatoire de Paris, Universit\'{e} PSL, Sorbonne Universit\'{e}, 
77 Avenue Denfert-Rochereau, 75014 Paris, France}

\date{Accepted . Received ; in original form }

\pubyear{2022}

\begin{document}
\label{firstpage}
\pagerange{\pageref{firstpage}--\pageref{lastpage}}
\maketitle
\begin{abstract}
Apart from being chaotic, the inner planets in the Solar System constitute an open system, 
as they are forced by the regular long-term motion of the outer ones. No integrals of motion can bound a priori the stochastic wanderings in their high-dimensional phase space. 
Still, the probability of a dynamical instability is remarkably low over the next 5 billion years, 
a timescale thousand times longer than the Lyapunov time. The dynamical half-life of Mercury has indeed been estimated 
recently at 40 billion years. By means of the computer algebra system TRIP, we consider a set of dynamical models 
resulting from truncation of the forced secular dynamics recently proposed for the inner planets at different degrees 
in eccentricities and inclinations. Through ensembles of $10^3$ to $10^5$ numerical integrations spanning 5 to 100 Gyr, 
we find that the Hamiltonian truncated at degree 4 practically does not allow any instability over 5 Gyr. 
The destabilisation is mainly due to terms of degree 6. This surprising result suggests an analogy 
to the Fermi-Pasta-Ulam-Tsingou problem, in which tangency to Toda Hamiltonian 
explains the very long timescale of thermalisation, which Fermi unsuccessfully looked for. 
\end{abstract}



\section{Introduction}
Even though the planet orbits in the inner Solar System (ISS) are chaotic with a Lyapunov time 
of about 5 million years \citep{laskar1989,laskar1990,sussman1992,Mogavero2021}, they are still statistically 
very stable over a timescale that is a thousand times longer. 
The probability of a Mercury eccentricity higher than 0.7 over the next 5 billion years, for example, is about 1\% from 
direct integrations of the Solar System \citep{Laskar2009,abbot2021}. This percentage agrees with the statistics 
of a dynamical instability observed in secular models where the dynamics is averaged over the planet mean longitudes \citep{laskar2008,Mogavero2021}. 
The statistical stability of the ISS over the remaining lifetime of the Sun as a main sequence star is intriguing, 
if one considers that it represents an open system, as it is forced by the very regular motion of 
the outer planets \citep{laskar1990,Mogavero2021}. No exactly conserved quantities, such as the energy or angular momentum, 
can bound a priori the chaotic wanderings of the system in its high-dimensional phase space. 

The disproportion between the Lyapunov time and the destabilisation timescale of the ISS has been addressed 
by \citet{batygin2015}, building on previous works by \citet{lithwick2011} and \citet{boue2012}. 
\citet{boue2012} consider the first-order 
secular dynamics of a mass-less Mercury in the gravitational field of all the other planets, 
whose orbits are predetermined to a quasi-periodic form. They use a multipolar 
expansion of the Hamiltonian to show that very high Mercury eccentricities appear in the reduced 
phase space of the resonance $g_1-g_5$ (involving the fundamental precession frequencies of the Mercury and 
Jupiter perihelia), which confirms the role of this harmonic in the destabilisation of the ISS 
\citep{laskar2008,Batygin2008,Laskar2009}. \cite{batygin2015} expand the secular Hamiltonian to degree 4 in 
eccentricities and inclinations of the planets, and study a few of its Fourier harmonics related to the
fundamental frequencies $g_1$, $g_2$, $g_5$, $s_1$, and $s_2$. Their simplified dynamics is however much 
more unstable than realistic models, the typical time for the destabilisation of Mercury orbit being 
around 1 Gyr \citep{woillez2020}. 
Recently, \citet[\ML{} from now on]{Mogavero2021} have proposed the model of a forced secular ISS, in which the outer 
planets only are frozen to quasi-periodic orbits. With a numerical experiment over 100 Gyr, 
they estimate the dynamical half-life of Mercury at 40 Gyr, consistently with the small probability of an 
instability over 5 Gyr. 

Here we employ the computer algebra software TRIP \citep{gastineau2011trip, TRIP} to perform truncation 
of the forced secular ISS at different degrees in eccentricities and inclinations. 
Through ensembles of $10^3$ to $10^5$ numerical integrations spanning 5 to 100 Gyr, we show how dynamical 
contributions usually deemed as unimportant, that is, high-degree terms of the Hamiltonian and non-resonant harmonics, 
strongly affect the probability of an instability over 5 Gyr. 

\begin{table*}
\centering
\begin{tabular}{c c c c c c c c c c} 
\hline\hline 
\rule{0pt}{1em}
$e_\textrm{max}$  & $\Hiss_{4}$ & $\Hiss_{6}$ & $\Hiss_{8}$ & $\Hiss_{10}$ & \multicolumn{2}{c}{$\Hiss$} & $\mathcal{L}_4$ & $\mathcal{L}_6$ & LG09 \\ 
\hline
\rule{0pt}{1em}
\input tables/table_e1m_5Gyr.out
\hline  
\end{tabular}
\caption{Probability $P(\sup_{t \leq  5 \, \textrm{Gyr}}e_1(t) \geq e_\textrm{max})$ in percent and its 90\% confidence interval, where $e_1$ is Mercury's eccentricity, 
for the dynamical models $\Hiss_{2n}$, $\Hiss$, $\mathcal{L}_{2n}$, and LG09. 
LG09 represents the 2\,501 direct integrations of \citet{Laskar2009}. $\Hiss$ denotes the 10\,560 orbital solutions of Gauss' dynamics in \ML{}, 
and the two values of the last two rows of $\Hiss$ correspond to the lower and upper bounds of the estimations, as explained in the text.}
\label{tab:5Gyr}
\end{table*}

\section{Dynamical models}
In the forced secular model of the ISS (detailed presentation in \ML{}), the orbits of the outer planets are 
predetermined to a quasi-periodic form, whose frequencies and amplitudes are inferred from frequency analysis 
\citep{laskar1988, Laskar2005} of a comprehensive model of the Solar System \citep{Laskar2011}. 
The secular gravitational interactions are considered at first order in planetary masses, which corresponds 
to Gauss' dynamics of Keplerian rings \citep{Gauss1818}, and the leading contribution of general relativity (GR) 
is included. We remark that the shift of the frequency $g_1$ due to GR is comparable to the half-width of the principal 
secular resonances \citep{Mogavero2022}, and cannot be considered as a small correction. 

With the aid of TRIP, the secular Hamiltonian $\Hsec$ of the entire Solar System, at first order in planetary masses, can be expanded in series 
of the complex \Poincare{} variables of the planets, i.e. $(x_i, \bar{x}_i, y_i, \bar{y}_i)_{i=1}^8$ \citep{laskar1995}. The planets are indexed 
in order of increasing semi-major axis, as usual. 
Truncation at total degree $2n$ results in a polynomial Hamiltonian $\Hsec_{2n}$. When the predetermined orbits of 
the outer planets $(x_i(t), y_i(t))_{i=5}^8$ are substituted, one obtains the Hamiltonian of the forced ISS truncated 
at degree $2n$, i.e. $\Hiss_{2n}((x_i, y_i)_{i=1}^4, t) = \Hsec_{2n}((x_i, y_i)_{i=1}^4, (x_i=x_i(t), y_i=y_i(t))_{i=5}^8)$. 
The non truncated Hamiltonian, formally $\Hiss = \Hiss_\infty$, represents Gauss' dynamics of the forced ISS. 

At the lowest degree, $\Hiss_{2}$ describes an integrable forced Laplace-Lagrange dynamics. 
Its analytical solution can be obtained by a canonical transformation to the complex proper modes 
variables $(u_i, v_i)_{i=1}^4$, with corresponding action-angle variables $(\Chi_i,\chi_i; \Psi_i,\psi_i)$ such that 
$(u_i = \sqrt{\Chi_i} \E^{-j \chi_i}; v_i = \sqrt{\Psi_i} \E^{-j \psi_i})$\footnote{$\E$ represents the 
exponential operator, $j$ stands for the imaginary unit.}. When expressed in these action-angle variables, 
the truncated Hamiltonian is a finite Fourier series: 
\begin{equation}
\label{eq:hamiltonian}
\begin{aligned}
\Huv_{2n}(\vecI,\vectheta,t) = 
\sum_{\veck, \vecl}
\fourcoeff{2n}{k}{\ell} (\vecI)
\E^{j \left( \vec{k} \cdot \vec{\theta} + \vec{\ell} \cdot \out{\vec{\omega}} t \right)}, \quad 
\fourcoeff{2n}{k}{\ell} = \sum_{p=1}^n \fourcoeff{(2p)}{k}{\ell}, 
\end{aligned}
\end{equation}
where $\vecI = (\vecChi,\vecPsi)$ and $\vectheta = (\vecchi,\vecpsi)$ are the eight-dimensional vectors of 
the action and angle variables, respectively, $t$ is the time, $\out{\vec{\omega}} = (g_5,g_6,g_7,g_8,s_6,s_7,s_8)$ is the 
septuple of the constant fundamental frequencies of the outer orbits \citep{laskar1990}, 
and $(\veck, \vecl) \in \mathbb{Z}^8 \times \mathbb{Z}^7 $ is the wave vector of a given harmonic. 
The amplitude of a harmonic $\fourcoeff{2n}{k}{\ell}$ consists of partial contributions 
$\fourcoeff{(2p)}{k}{\ell}$ from terms of the same degree $2p \leq 2n$. To identify these partial contributions 
we define 
\begin{equation}
\mathcal{F}_{(2p)}^{\veck, \vecl} = 
\fourcoeff{(2p)}{k}{\ell} \E^{j \left( \vec{k} \cdot \vec{\theta} + \vec{\ell} \cdot \out{\vec{\omega}} t \right)}.
\end{equation}
The order of a harmonic is defined as the even integer $\| (\veck, \vecl) \|_1 \leq 2n$, where $\| \cdot \|_1$ denotes the 1-norm. 
Since the quasi-periodic form of the outer orbits contains harmonics of order higher than one, the dynamics of 
$\Hiss_{2n}$ and $\Huv_{2n}$ are not exactly the same. Yet, the difference is unimportant for the results of this work, 
so we shall treat the two Hamiltonians as equivalent from now on. 

\paragraph*{Second order in planetary masses.}
To investigate the effect of the order of the secular averaging on the long-term statistics, we employ the autonomous polynomial equations 
of motion of \citet{laskar1985,laskar1990} for the ensemble of the Solar System planets. These equations formally derive from a Hamiltonian of 
order two in masses and degree 6 in eccentricities and inclinations, and will be denoted as $\mathcal{L}_6$ throughout the paper. 
In this work, we also implement a variant of this dynamics, in which the equations for the inner planets are truncated at total degree 3 
in eccentricities and inclinations, while those of the outer planets are kept at degree 5 (Appendix \ref{app:L46}). This new model, denoted as 
$\mathcal{L}_4$, is meant as an analogue of $\Hiss_4$ at second order in masses. 

\section{Numerical Experiments} 
\label{sec:main}
We systematically derive the equations of motion for the truncated Hamiltonians $\Hiss_{2n}$ in TRIP. 
They are numerically integrated via an Adams PECE scheme of order 12, with a time step of 250 years. 
Typical integration times are given in \ML{} (table 1). 
\begin{figure}
\includegraphics[width=\columnwidth]{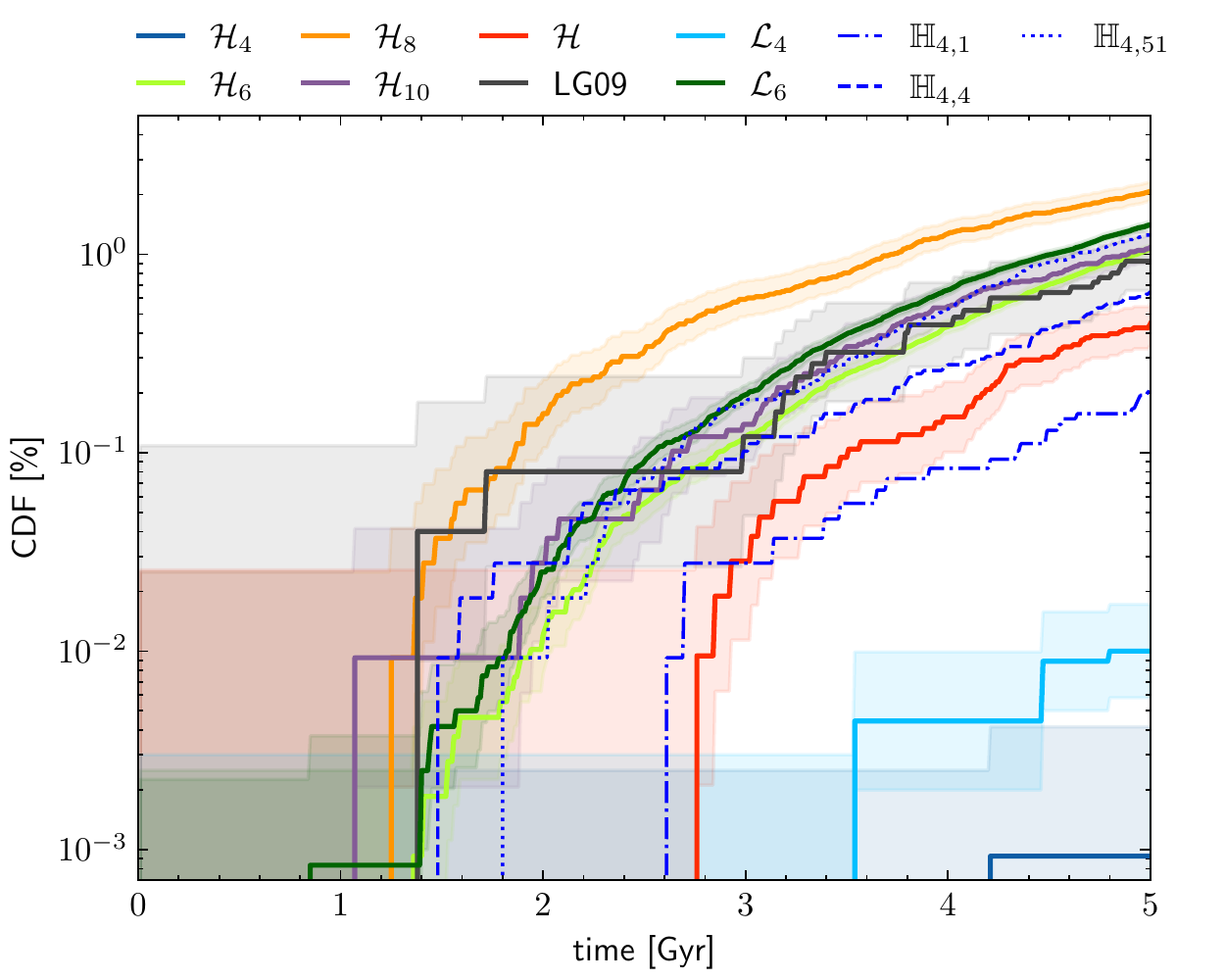}
\caption[cc]{CDF of the first hitting time of Mercury's eccentricity at 0.7 over 5 Gyr with 90\% piecewise confidence interval, for the dynamics 
$\Hiss_{2n}$, $\mathbb{H}_{4,m}$, $\Hiss$, $\mathcal{L}_{2n}$, and LG09. LG09 represents 2\,492 direct integrations \protect\footnotemark \citep{Laskar2009}, $\Hiss$ denotes 10\,560 solutions of Gauss' dynamics (\ML{}).}
\label{fig:5Gyr}
\end{figure}
\footnotetext{9 out of the original 2501 solutions were damaged during data storage.}
All the orbital solutions of $\Hiss_{2n}$ in this paper correspond to initial conditions taken from a unique ensemble of 108\,000 values very close to each other, and distributed according to:
\begin{equation}
\label{eq:IC}
x_i = x_i^\ast + \sigma \left( \operatorname{Re}\{x_i^\ast\} \, z_i + j \operatorname{Im}\{x_i^\ast\} \, z_i^\prime \right),
\end{equation}
where $x_i^\ast$ represents the nominal initial conditions for $\Hiss$ given in \ML{} (appendix D), 
$z_i, z_i^\prime \sim \mathcal{N}(0,1)$ are standard normal deviates, and $\sigma = 10^{-9}$.
An analogous expression holds for the variables $(y_i)$. Initial conditions for $(u_i,v_i)$ are directly derived from 
the transformation $(x_i,y_i) \rightarrow (u_i,v_i)$. 
For the first few million years, all the solutions reproduce the comprehensive direct simulation LaX13b (\ML{}), while they diverge from each other after about 100 Myr  
due to chaos. 
The choice of the initial distribution has an impact on the secular solutions that decreases with time because of chaotic diffusion \citep{hoang2021}. Therefore, the long-term statistics we present should not depend on its particular shape, but should rather reflect the nature of the dynamical models employed.

\begin{figure}
 \includegraphics[width=\columnwidth]{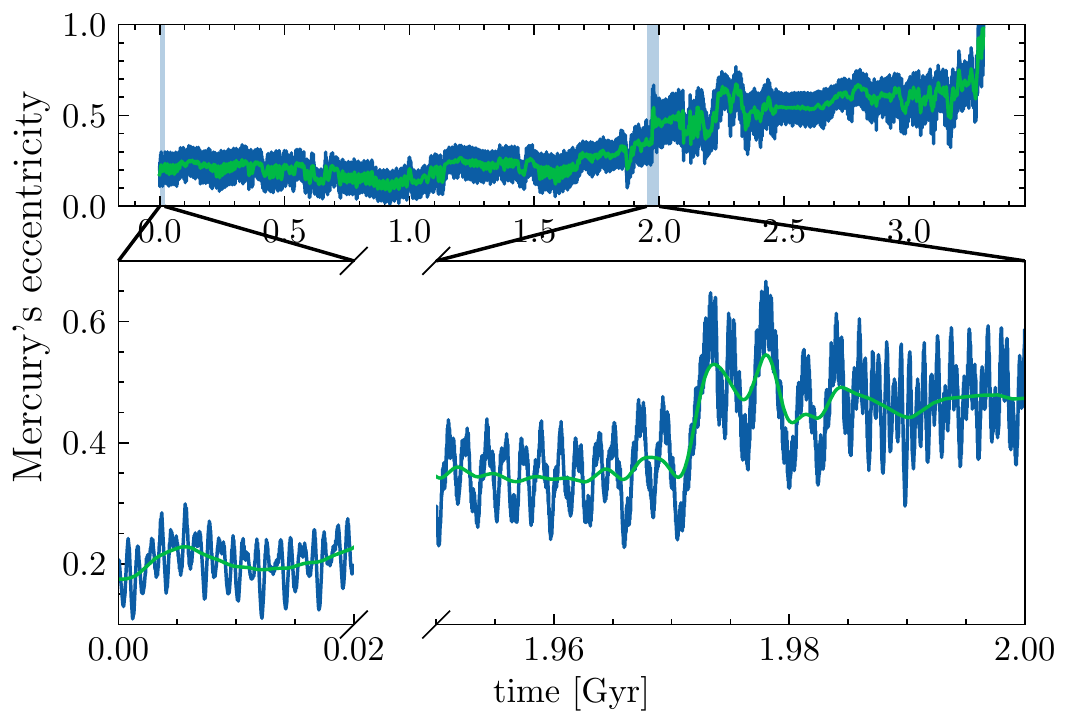}
 \caption{Temporal evolution of Mercury's eccentricity for an unstable solution of $\Hiss_6$ (blue curve) and its KZ-filtered value with 3 iterations of the moving average and a cutoff frequency of (5 Myr$)^{-1}$ (green curve). The initial period of 20 Myr and the period of the first activation of the resonance $g_1-g_5$ from 1.95 Gyr to 2 Gyr are enlarged in the lower panel. }
 \label{fig:e1}
\end{figure}

We compute 108\,000 solutions spanning 5 Gyr in the future for $\Hiss_{4}$ and $\Hiss_{6}$, and 10\,800 solutions for $\Hiss_{8}$ and $\Hiss_{10}$ over the same time interval. 
For each Hamiltonian, we prolong 1\,080 solutions to 100 Gyr. 
The statistics of $\mathcal{L}_6$ was first described in \citep{laskar2008} with 478 solutions integrated up to 5 Gyr. In this paper, we compute a much larger ensemble of solutions: 
120\,000 and 40\,000 solutions at degree 6 lasting for 5 Gyr and 100 Gyr in the future, respectively; 90\,000 and 10\,000 solutions at degree 4 ($\mathcal{L}_4$) spanning the same intervals. 
The statistics of this paper will be compared with those from previous works: the forced secular ISS without truncation in eccentricities nor in inclinations, i.e. Gauss' dynamics, 
denoted as $\Hiss$ (\ML{}); the direct integrations of the Solar System of \citet{Laskar2009} denoted as LG09. 

For each ensemble of solutions, we retrieve the statistics of the maximum value reached by the eccentricity of Mercury over a given timespan \citep{laskar1994}
This choice is motivated by the fact that the excitation of Mercury's eccentricity due to the resonance $g_1 - g_5$ is a precursor of the dynamical instability. 
Mercury's eccentricity at 5 Gyr typically ranges from 0 and 0.5 \citep[\ML]{laskar2008,Laskar2009}. 
The rare activation of the resonance $g_1-g_5$ allows a net transfer of angular momentum deficit \citep{laskar1997} from the outer planets to the ISS, 
and pump the eccentricity of Mercury to a higher value. Once the eccentricity of Mercury exceeds 0.7, 
the solutions enter an unstable regime, where close encounters and collisions involving Mercury become possible. 
Therefore, a Mercury's eccentricity higher than 0.7 shall be taken as a synonym of instability for the rest of the paper. 

All the secular solutions are stopped at numerical instability, except those of Gauss' dynamics which end at a secular collision, 
that is, the geometric intersection of the Keplerian ellipses of two planets (\ML{}). To have a more accurate comparison, we assume that after a secular collision, 
the maximum Mercury eccentricity of a Gauss' solution exceeds 0.9 shortly, which corresponds to the upper bounds of the column $\Hiss$ in Table \ref{tab:5Gyr}. This assumption for the solutions of $\Hiss$ is used for the remainder of the paper. In contrast, the lower bounds assume that the maximum eccentricity of Mercury of such solutions does not reach higher values after a secular collision, and correspond to the statistics reported in \ML{} (table 4). 

\begin{table}
\centering
\begin{tabular}{c  r r  r r} 
\hline
\hline 

\rule{0pt}{1em}

$i$ & Harmonic $\mathcal{F}_{(4)}^i$ & $\mathcal{C}^{\vec{k},\vec{\ell}}_{(4)} $
    & Harmonic $\mathcal{F}_{(6)}^i$ & $\mathcal{C}^{\vec{k},\vec{\ell}}_{(6)} $   \\
\hline
\input tables/ranking_harms_sol29.out
\hline
\end{tabular}
\caption{Rankings of Fourier harmonics. Partial contributions to $g_1$ (arcsec yr$^{-1}$) from the harmonics at degree 4 ($\mathcal{F}_{(4)}^i$) and 6 ($\mathcal{F}_{(6)}^i$), 
along the unstable solution of $\Hiss_6$ of Fig.~\ref{fig:e1}. The maximum filtered contributions are denoted by 
$\mathcal{C}^{\vec{k},\vec{\ell}}_{(2p)} = g_{1(2p)}^{\vec{k},\vec{\ell}}(t^\star)$, with $t^\star = \argmax_{t \leq T} |g_{1(2p)}^{\vec{k},\vec{\ell}}(t)|$ 
(Eqs.~\eqref{eq:hrmnc}, \eqref{eq:g1_filter}). For each partial degree, the harmonics are ranked according to $|\mathcal{C}^{\vec{k},\vec{\ell}}_{(2p)}|$ 
with $T=2\text{ Gyr}$, which is shortly after the first activation of the resonance $g_1-g_5$.} 
\label{tab:ranking}
\end{table}

\section{Statistics of Mercury's eccentricity}

\subsection{Small changes, big differences over 5 Gyr}

\begin{figure*}
\includegraphics[width=\textwidth]{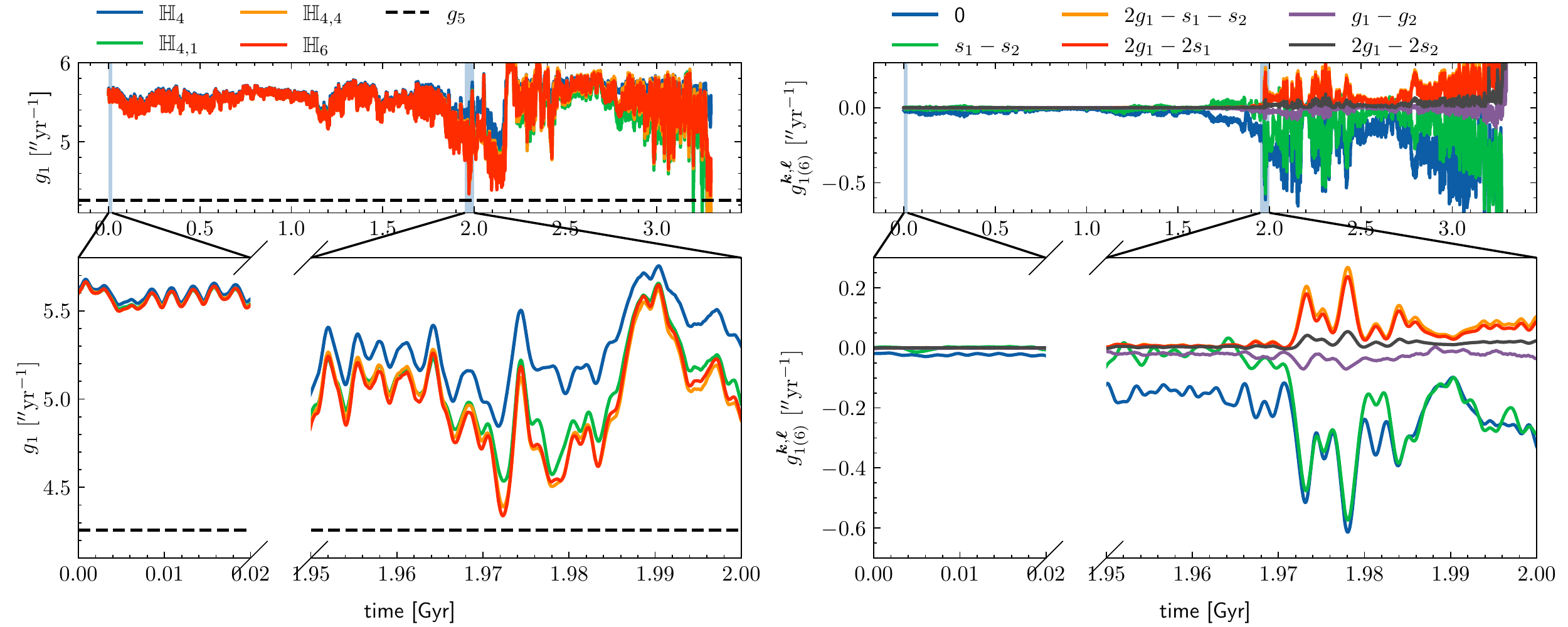}
\begin{minipage}{0.49\textwidth}
\subcaption{\label{fig:g1a} }
\end{minipage}
\begin{minipage}{0.49\textwidth}
\subcaption{\label{fig:g1b} }
\end{minipage}
 \caption{Temporal evolution of the filtered frequency $g_1$ defined from different Hamiltonians (left) and partial contributions $g_{1(6)}^{\vec{k},\vec{\ell}}$
 at degree 6 from the six leading harmonics of Table~\ref{tab:ranking} (right) along the unstable integration of $\Hiss_6$ of Fig.~\ref{fig:e1}. 
 The low-pass filter has a cutoff frequency of (1 Myr$)^{-1}$. The initial period of 20 Myr and the period of the first activation of the resonance $g_1-g_5$ 
 from 1.95 Gyr to 2 Gyr are enlarged in the lower panels.}
 \label{fig:g1}
\end{figure*}

Table~\ref{tab:5Gyr} shows for each dynamical model the percentages of solutions whose Mercury's maximum eccentricity over 5 Gyr 
reaches various values, from 0.35 to 0.9. We report statistical confidence bounds estimated by \citet{wilson1927}'s score interval at 90\% level.
A temporal evolution of the statistics is presented in Figure \ref{fig:5Gyr}, which displays the cumulative distribution functions (CDFs) 
of $\tau = \inf_{t} \{ e_1(t) \geq 0.7 \}$, that is, the first time that the eccentricity of Mercury $e_1$ reaches the threshold of 0.7 
along a given solution (the variation of the CDFs with different thresholds is studied in Appendix \ref{app:thresholds}). 
The values of the curves at 5 Gyr coincide with the line of 0.7 of Table \ref{tab:5Gyr}.

The most striking results from Figure \ref{fig:5Gyr} and Table \ref{tab:5Gyr} lie in the statistics of the models of degree 4, $\Hiss_4$ and $\mathcal{L}_4$. 
The probability of a high Mercury eccentricity over 5 Gyr is around $1\%$ in LG09, which is considered as the reference model, and this is reproduced 
up to a factor of two by all the models of degree 6 and higher. Nevertheless, the dynamics of $\Hiss_{4}$ is much more stable, with only one solution 
among 108\,000 in which Mercury's eccentricity exceeds 0.7, for an estimated probability of $10^{-5}$, a thousand times smaller than that of the reference model. 
At second order in planetary masses, the disparity between $\mathcal{L}_4$ and $\mathcal{L}_6$ is two orders of magnitude, which 
is still substantial. The CDF of $\mathcal{L}_6$ is slightly greater than that of $\mathcal{H}_6$, which shows that the contribution of the second order in planetary masses 
is small and destabilizing. Nevertheless, for such a stable model like $\mathcal{H}_4$, the second order can still raise the instability rate by one order of magnitude. 

The great stability of the $\Hiss_4$ dynamics shows that the low probability of 1\% for an instability of Mercury orbit over 5 Gyr should be interpreted as a perturbative effect, with the leading 
contribution coming from the Hamiltonian terms of degree 6. The practical stability of $\Hiss_4$ over 5 Gyr is unexpected, since 
it still reproduces the chaotic dynamics of the ISS with the same long-term statistical distribution of the maximum Lyapunov exponent 
as in $\Hiss_6$ or Gauss' dynamics \citep{Mogavero2022}. It also shows the same destabilisation mechanism, that is, the activation of the resonance $g_1-g_5$. 
Previous works on the instability of Mercury orbit studied a simplified dynamics in which only a few Fourier harmonics of $\Hiss_{4}$ are considered \citep{batygin2015,woillez2020}. 
This latter model is actually considerably more unstable than the reference model LG09, in deep contrast with the practical stable dynamics of $\Hiss_{4}$ over 5 Gyr. 

The secular models of degree higher than 4 give predictions that generally agree with the $N$-body integrations LG09. While Gauss's dynamics tends to underestimate, 
and $\Hiss_8$ overestimates the instability probability, $\Hiss_{10}$ gives accurate predictions (this is probably a coincidence related to the behaviour of the series 
expansion). At times shorter than 5 Gyr and for a threshold of 0.7, the estimations of $\Hiss_6$, $\Hiss_{10}$, and $\mathcal{L}_6$ generally agrees with the results 
of a refined method of rare event detection applied to direct integrations \citep{abbot2021}\footnote{The definition of instability in \citet{abbot2021} is however different.}: for $\mathcal{L}_6$, 
the probability of Mercury orbit having eccentricity larger than 0.7 in the next 2 Gyr is $0.025\%$ with a $90\%$ confidence interval ($0.019 \%, 0.034 \%$), while
it is $0.2\%$ with a $90\%$ confidence interval ($0.18 \%, 0.22 \%$) over 3 Gyr. 
The Hamiltonians of degree 4 and 6 show a relatively pronounced decay of the probability from 0.7 to 0.9 (see also Fig.~\ref{fig:CDFs_compare}). This interesting phenomenon 
is inherent to degree 6 or lower, because the probability of the models of higher degree is fairly constant across high values of Mercury's eccentricity: 
if $e_1$ reaches 0.7 along a solution, it also probably goes beyond 0.9. 

\subsection{Ranking of harmonics according to their contributions to \texorpdfstring{$g_1$}{g\_1}}
\label{sec:rank}
In order to explain the difference between the statistics of $\Hiss_{4}$ and $\Hiss_{6}$, we shall switch to the proper mode variables and the Fourier representation 
of Eq.~\eqref{eq:hamiltonian}. The Hamiltonian $\Huv_{6}$ contains substantially more harmonics than $\Huv_{4}$, 69\,339 compared to 2\,748. For each harmonic of $\Huv_{4}$, 
$\Huv_{6}$ includes additional terms of degree 6 in its amplitude. Despite the large difference in the number of terms, most of the contributions of $\Huv_{6}$ are negligible. 
We aim to identify here the Fourier harmonics that have an important impact on the destabilisation mechanism, that is, the activation of the resonance $g_1-g_5$. 
Because $g_5$ is constant in the forced dynamics, we shall focus on the fundamental precession frequency of Mercury perihelion $g_1$. 

Following \ML{}, the instantaneous value of the frequency $g_1$ for the Hamiltonian $\Huv_{2n}$ is defined as:
\begin{equation}
\label{eq:g1_instant}
\hat{g}_{1}^{(2n)} = -\dot{\theta}_1 = -\frac{\partial \Huv_{2n}}{\partial I_1} = 
\sum_{\vec{k},\vecl}
\sum_{p=1}^n \hat{g}_{1(2p)}^{\vec{k},\vec{\ell}},
\end{equation}
where the partial contribution at degree $2p$ of each harmonic is
\begin{equation} 
\label{eq:hrmnc}
\hat{g}_{1(2p)}^{\vec{k},\vec{\ell}} = - \frac{\partial \mathcal{F}_{(2p)}^{\veck, \vecl} (\vecI) }{\partial I_1}  = - \frac{\partial \fourcoeff{(2p)}{k}{\ell} (\vecI)}{\partial I_1}
\E^{j \left( \vec{k} \cdot \vec{\theta} + \vec{\ell} \cdot \out{\vec{\omega}} t \right)}. 
\end{equation}
In this form, each harmonic manifests its importance via its direct contribution to $g_1$, which varies along an orbital solution according to the position in the phase space, i.e. $\hat{g}_{1(2p)}^{\vec{k},\vec{\ell}}(t) = \hat{g}_{1(2p)}^{\vec{k},\vec{\ell}}(\vecI(t),\vectheta(t),t)$. 
To identify the main harmonics involved in the destabilisation of the dynamics, 
Eqs. \eqref{eq:g1_instant} and \eqref{eq:hrmnc} are evaluated along unstable solutions.
Short-term oscillations are suppressed by the low-pass Kolmogorov-Zurbenko (KZ) filter \citep{Zurbenko2010}, 
which is applied to the instantaneous frequency $g_1$ and its harmonic contributions. 
We use the KZ filter with 3 iterations of the moving average and a cutoff frequency of (1 $\text{Myr})^{-1}$ (\ML, appendix B) to obtain the filtered values 
\begin{equation}
\label{eq:g1_filter}
g_{1}^{(2n)} = \text{KZ} (\hat{g}_{1}^{(2n)}), \quad   
g_{1(2p)}^{\vec{k},\vec{\ell}} = \textnormal{KZ} (\hat{g}_{1(2p)}^{\vec{k},\vec{\ell}})
\end{equation}
The harmonics can then be ranked according to the maximum value of their absolute filtered contribution over the time interval $[0,T]$. 
The timespan $T$ is chosen to be slightly larger than the time of the first activation\footnote{Throughout the paper, by \emph{activation} we mean the exploration of the chaotic zone 
of the resonance, independently of the entrance in a libration state.} of the resonance $g_1-g_5$. 
After this point, the system either exhibits a secular collision right away 
or enters a period of excited dynamics before an eventual collision. This unstable state typically lasts longer for a solution of $\Hiss_6$ than for a Hamiltonian 
of higher degree.

We establish the harmonic ranking on an unstable solution of $\Hiss_6$, whose Mercury's eccentricity over time is shown in Fig.~\ref{fig:e1} 
(the ranking of the leading harmonics is quite robust when we switch to other unstable solutions). 
The first entrance into the chaotic zone of the resonance $g_1-g_5$ occurs just after 1.97 Gyr (see Fig.~\ref{fig:g1a}), during which the eccentricity of Mercury 
is pumped to 0.65 and the harmonic contributions generally reach their maximum values (see Fig.~\ref{fig:g1b}). The ranking is computed over the first 2 Gyr to capture 
the contributions of the harmonics at the resonance. Table~\ref{tab:ranking} shows two harmonic rankings based on the partial contributions at degree 4 and 6, respectively. 
It is surprising to find that the contributions to $g_1$ at degree 6 are slightly less, but still roughly the same amount as those at degree 4.
Because the principal contributions at degree 6 come from harmonics of order 2 and 4, what $\Huv_6$ mainly offers is not new resonances, but rather corrections to the existing harmonics of $\Huv_4$.
The corrections at degree 6 help to push $g_1$ toward $g_5$ and bring the solution closer to the destabilizing resonance.
Geometrically speaking, in the phase space the resonance $g_1-g_5$ defined by $\Huv_6$ is closer to the current ISS than that of $\Huv_4$. 

Figure~\ref{fig:g1b} gives a closer look at the time evolution of the leading harmonic contributions to $g_1$ at degree 6.
They are small at the beginning when the solution is stable, but get much bigger when the eccentricity of Mercury becomes higher, 
that is during and after the first activation of the resonance $g_1-g_5$ at 1.97 Gyr. During this period, which is shown in the 
lower panel of Fig.~\ref{fig:g1b}, the strongest terms are the null-frequency harmonic, i.e. the integrable part of the 
Hamiltonian $\fourcoeff{(6)}{0}{0}$, and the harmonic $s_1-s_2$, which also enters resonance.
These two terms tend to destabilize the system by decreasing $g_1$ by substantial amounts, which are even greater than 
the leading GR correction of 0.4\arcsec{} yr$^{-1}$ at degree 2 at some point. In the opposite direction, the two harmonics $2g_1-(s_1+s_2)$ and $2(g_1-s_1)$ raise 
$g_1$, moving it away from $g_5$. Although these terms are non resonant, they are extremely crucial for the stability of Mercury orbit (see Section~\ref{unstable}).
Other harmonics also contribute to $g_1$ at degree 6 in an alternating pattern, but to a lesser extent. 

To confirm the crucial role of the terms of degree 6, we add them to $\Huv_4$ to construct partial Hamiltonians \citep{Mogavero2022}:
\begin{equation}
\label{eq:partial_ham}
\Huv_{4,m} = \Huv_4  + \sum_{i=1}^m  \mathcal{F}_{(6)}^i,
\end{equation}
where $\mathcal{F}_{(6)}^i = \mathcal{F}_{(6)}^{\veck_i,\vecl_i}$ 
is the $i$th harmonic from the ranking at degree 6 of Table~\ref{tab:ranking}, and $m$ is the total number of such harmonics that are considered. 
Figure~\ref{fig:g1a} shows the filtered $g_1$ computed from different Hamiltonians along the same unstable trajectory of $\Hiss_6$ of Fig.~\ref{fig:e1}. 
Initially, when the solution is stable and Mercury's eccentricity is relatively low, the frequency $g_1^{(4)}$ of $\Huv_4$ is almost indistinguishable 
from the corresponding $g_1^{(6)}$ of $\Huv_6$. Across the activation of the resonance $g_1-g_5$, the difference between the two frequencies 
becomes considerable: $g_1^{(6)}$ almost reaches $g_5$, while $g_1^{(4)}$ does not. The difference is mainly due to the integrable term $\fourcoeff{(6)}{0}{0}$, 
which is included in $\Huv_{4,1}$, and to the first leading harmonics contained in $\Huv_{4,4}$.

The statistics of the high Mercury eccentricities from $\Huv_{4,m}$ should approximate that of $\Huv_6$ better than $\Huv_4$. In order to test this expectation, 
we integrate the dynamics of $\Huv_{4,1}$, $\Huv_{4,4}$ and $\Huv_{4,51}$ from 10\,800 and 1\,080 initial conditions over 5 Gyr and 100 Gyr, 
respectively. The initial conditions are taken from the same ensembles employed for $\Hiss_{2n}$. 
The CDFs of the first time that Mercury eccentricity reaches 0.7 for $\Huv_{4,m}$ are shown in Fig.~\ref{fig:5Gyr}. 
The wide discrepancy between $\Hiss_4$ and $\Hiss_6$ is first bridged by adding the integrable term $\fourcoeff{(6)}{0}{0}$, with which the curve of $\Huv_{4,1}$ 
attains a probability of $0.2\%$ at 5 Gyr. Including the next three leading harmonics brings the curve to the same level as Gauss' dynamics. 
Adding additional terms makes the statistics oscillate around that of $\Huv_6$. 

The impact of the choice of the initial conditions on the present analysis deserves a discussion. As stated in Section~\ref{sec:main}, the nominal initial conditions 
of the truncated forced dynamics $\Hiss_{2n}$ are chosen to be the same as those of Gauss' dynamics $\Hiss$. In principle, they should be adapted 
to each model according to the harmonics that are dropped from the full Hamiltonian \citep[][\ML{}]{LaskarSimon1988}. Nevertheless, the lack of adjustment of the 
nominal initial conditions has a negligible effect in our study. First of all, the harmonic contributions to $g_1$ in Table~\ref{tab:ranking} are established on an 
orbital solution of $\Hiss_6$: the change in the initial conditions with respect to $\Hiss$ is of only degree 8 in eccentricities and inclinations of the planets \citep{Morbidelli2002}. 
Secondly, $\Hiss_4$ and all the partial Hamiltonians considered in Eq.~\eqref{eq:partial_ham} contain the entire contribution from terms 
of degree 4. Therefore, the change in the initial conditions is still of degree 6. These considerations indicate that all the models considered here reproduce consistently 
the dynamics of the ISS on short (secular) timescales, as shown for the frequency $g_1$ in the lower panel of Fig.~\ref{fig:g1a}.
Moreover, the impact on long-term statistics of small differences among ensembles of initial conditions generally decreases with time because of chaotic diffusion \citep{hoang2021}. 
As a result, our findings should not be sensitive to the initial displacement in the phase space, but rather reflect the distinctive long-term behaviour of the different models. 

\subsection{Importance of non-resonant harmonics} 
\label{unstable} 
\begin{figure}
 \includegraphics[width=\columnwidth]{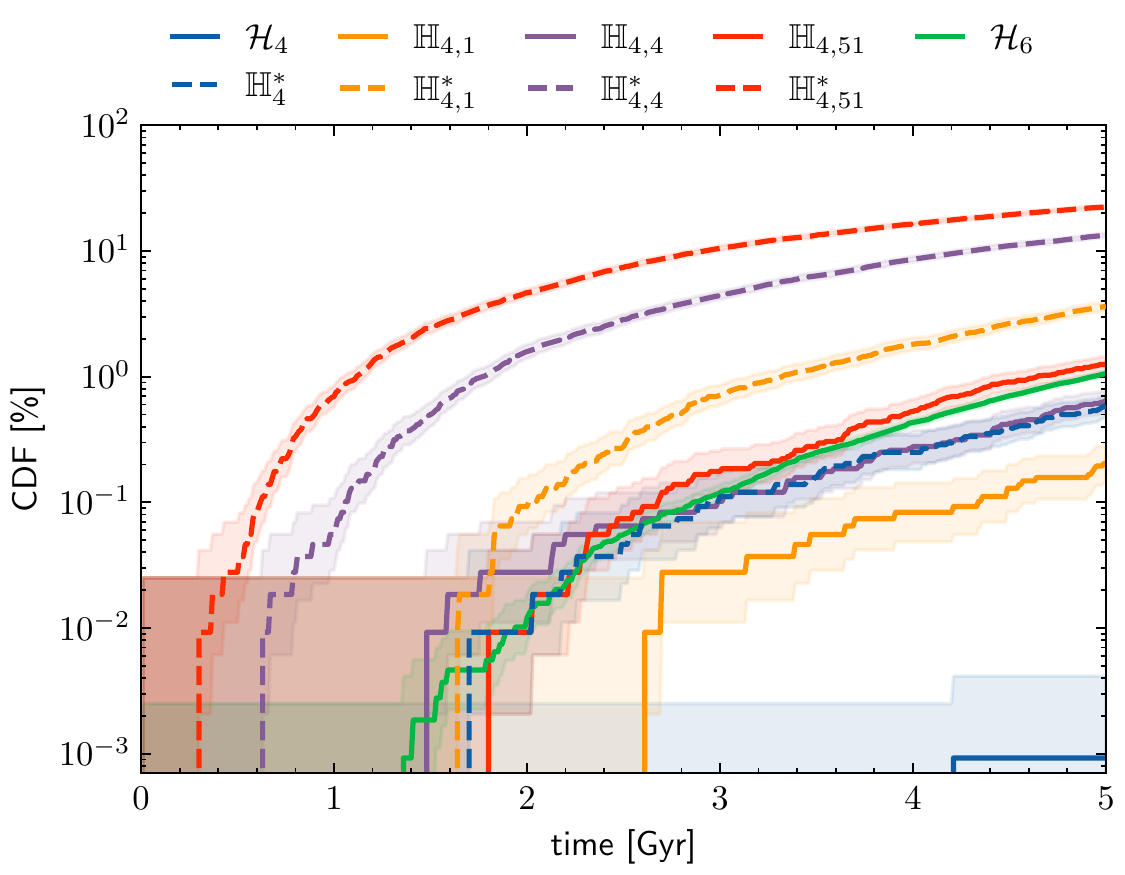}
 \caption[cc]{CDFs of the first hitting time of Mercury's eccentricity at 0.7 over 5 Gyr for the dynamical models $\Hiss_4$, $\Hiss_6$, 
 $\mathbb{H}_{4,m}$ (solid lines) and $\Huv_4^*$, $\mathbb{H}^*_{4,m}$ (dashed lines). The Hamiltonians $\Huv_4^*$ and $\mathbb{H}^*_{4,m}$ 
 exclude from $\Huv_4$ and $\mathbb{H}_{4,m}$, respectively, the entire contribution of the three non-resonant harmonics $2g_1-(s_1+s_2)$, 
 $2(g_1-s_1)$, and $2(g_1-s_2)$ of Table \ref{tab:ranking}. 
 } 
 \label{fig:CDF_unstable}
\end{figure}
We have shown the importance of harmonics at degree 6 by adding them to $\Huv_4$ to construct partial Hamiltonians. 
Among the leading terms, there are several non-resonant harmonics, which are often considered unimportant when constructing simplified models. 
Among the leading non-resonant harmonics of Table~\ref{tab:ranking}, we consider $2g_1-(s_1+s_2)$, $2(g_1-s_1)$, and $2(g_1-s_2)$, 
to highlight their role in stabilizing the ISS. We shall subtract the entire contribution of these three harmonics from the Hamiltonians $\Huv_{4}$ 
and $\Huv_{4,m}$, to define new partial Hamiltonians denoted as $\Huv^*_{4}$ and $\Huv_{4,m}^*$, respectively.
The values of $m$ are chosen to be the same as in Section~\ref{sec:rank}, that is, $m \in \{1, 4, 51\}$. 
We integrate the equations of motion defined by $\Huv^*_{4}$ and $\Huv_{4,m}^*$ over 5 Gyr from the same ensemble of initial conditions 
defined in Section~\ref{sec:main}, to obtain 10\,800 solutions.

Figure~\ref{fig:CDF_unstable} shows the comparison between $\Huv_4$, $\Huv_{4,m}$ and $\Huv^*_{4}$, $\Huv^*_{4,m}$ for the CDF of the first time 
that Mercury's eccentricity reaches 0.7 over 5 Gyr. For all the models, removing the three non-resonant harmonics makes the 
dynamics significantly more unstable, with at least one order of magnitude of difference. For comparison, the dynamics of $\Hiss_4$ 
is a thousand times more stable than $\Hiss_6$ over 5 Gyr, but taking away the three harmonics brings the model $\Huv_4^*$ basically to the 
same level of instability of $\Hiss_6$. If we consider the Hamiltonian $\Huv_{4,51}^*$, based on $\Huv_{4,51}$ which is the closest 
dynamics to $\Hiss_6$ among the presented partial Hamiltonians, its probability of instability is around $20\%$ at 5 Gyr, that is, 
twenty times more than the instability rate of $\Hiss_6$. 
These numerical experiments show the crucial role of these non-resonant harmonics in stabilizing the ISS. 
Interestingly enough, all the three harmonics permits the exchange of angular momentum deficit between the eccentricity and inclination
degrees of freedom, that is, between the proper modes $(u_i)$ and $(v_i)$. These results also show the sensitivity of the destabilisation 
probability to the details of the dynamics, and may explain, at least partially, the 
great instability shown by the simplified models considered in literature \citep{batygin2015,woillez2020}. 

\subsection{Statistics over 100 Gyr}
\begin{figure}
 \includegraphics[width=\columnwidth]{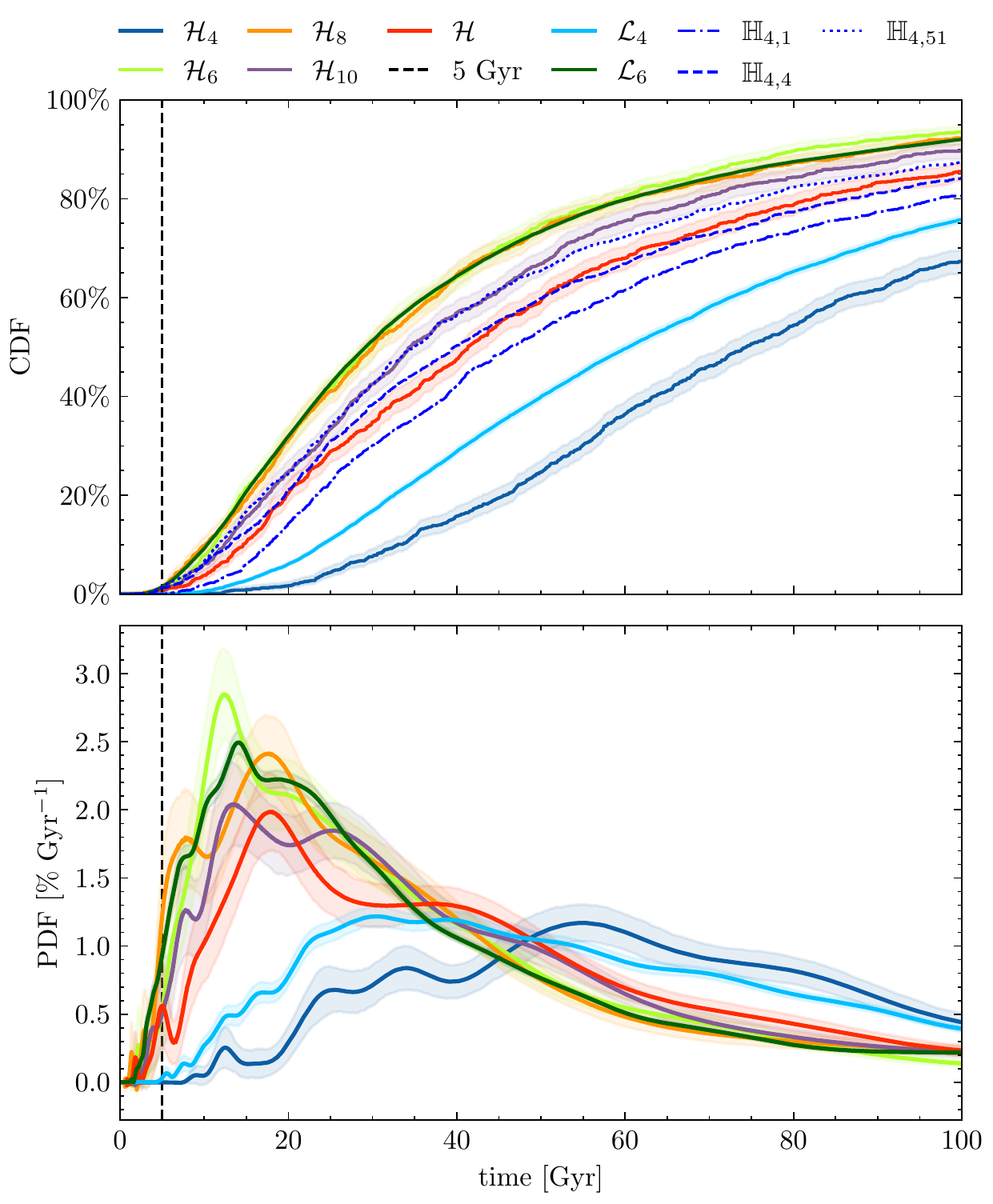}
 \caption{CDF and PDF of the first hitting time of Mercury's eccentricity at 0.7 over 100 Gyr with 90\% piecewise confidence interval, for the dynamical models $\Hiss_{2n}$, $\mathbb{H}_{4,m}$, $\mathcal{L}_{2n}$, and $\Hiss$. $\Hiss$ denotes the 1\,042 orbital solutions of Gauss' dynamics in \ML{}.} 
 \label{fig:100Gyr}
\end{figure}
To explore the dynamics in a regime where highly excited orbits no longer represent rare events, we follow \ML{} and prolong 1080 integrations 
of the different Hamiltonians previously considered to 100 Gyr. 
Figure~\ref{fig:100Gyr} shows the CDFs of the first time that Mercury eccentricity reaches 0.7 and the corresponding probability density functions (PDFs). The PDFs are estimated by the debiased 
kernel density estimation (KDE) method \citep{cheng2019}, with Gaussian kernel and \citet{silverman1986}'s rule-of-thumb bandwidth (Appendix~\ref{supp:PDF}). We use the log transformation 
and the pseudo-data method \citep{cowling1996} to remove the boundary effects induced by the KDE at 0 and 100 Gyr, respectively. The confidence intervals of the PDFs are estimated 
by bootstrap \citep{Efron1979} of the debiased KDEs; for the CDFs, we use Wilson's score interval. 

The CDFs of $\Hiss_6$ and $\Hiss_8$ are close to each other, with medians of 30 Gyr, while that of $\Hiss_{10}$ is around 35 Gyr. 
The increasing values of the medians may suggest a convergence toward the value of 40 Gyr of Gauss' dynamics.
On the other hand, the difference between $\Hiss_4$ and the other truncated forced dynamics is still considerable. 
The median time for $\Hiss_4$ is 75 Gyr, roughly doubling the value of $\Hiss_6$. 
If we assume that the PDFs follow a Levy distribution $\rho(\tau) = (T_0/\pi \tau^3)^{1/2} \E^{-T_0/\tau}$ over short times (\ML{}), 
with $T_0$ proportional to the median of the distribution, one easily understands how a difference by a factor of two in the medians of the PDFs 
results in very different probabilities over 5 Gyr. Indeed, the fact that the destabilisation over 5 Gyr is a rare event greatly amplifies the 
disparity between $\Hiss_4$ and the models of higher degree.

There is practically no difference between the statistics of $\Hiss_6$ and $\mathcal{L}_6$ over this timescale, which confirms the secondary effect of the second order 
in masses for the forced ISS and the statistics of the high Mercury eccentricities in particular. However, this effect is magnified for $\Hiss_4$, the CDF of $\mathcal{L}_4$ approaching halfway 
the curves of higher degrees, with a median time of around 60 Gyr. 
Figure~\ref{fig:100Gyr} also shows the CDFs of $\Huv_{4,m}$, highlighting the impact of the leading harmonics at degree 6. With only the integrable term 
$\fourcoeff{(6)}{0}{0}$ considered, the CDF of $\Huv_{4,1}$ is already close to that of $\Hiss$. When additional harmonics are added, their CDFs approach 
the curve of $\Hiss_6$, as shown by $\Huv_{4,4}$ and $\Huv_{4,51}$. 

\section{Discussion}
Our findings suggest a remarkable analogy between the secular ISS and the Fermi-Pasta-Ulam-Tsingou 
(FPUT) problem, which consists in a chain of coupled weakly-anharmonic oscillators \citep{Fermi1955}. 
This is basically the same kind of interactions as in the secular planetary problem. Differently from Fermi's expectations, 
the proper modes of oscillation of the FPUT dynamics remain far from the equipartition invoked in equilibrium statistical mechanics for a very long time. 
This has been related to the closeness of the FPUT problem to the integrable Toda dynamics, which does not allow 
any thermalisation of its action variables \citep{Henon1974,Flaschka1974,Ferguson1982,Benettin2013}. 
Although not integrable, and indeed chaotic, the Hamiltonian $\Hiss_4$ plays a role similar to the Toda Hamiltonian, 
as it does not allow essentially any dynamical instability over 5 Gyr. 
The main question at this point is why the dynamics of $\Hiss_4$ is practically stable over 5 Gyr. 
Once this is assessed, the small 1\% probability of an instability of the ISS may be conceived as a natural perturbative 
effect of terms of degree 6 and higher. 

\section*{Acknowledgements}
The authors are indebted to M.~Gastineau for his support with TRIP. 
N.~H.~H. is supported by a PhD scholarship of the CFM Foundation for Research.
F.~M. is supported by a grant of the French Agence Nationale de la Recherche 
(AstroMeso ANR-19-CE31-0002-01) and has been supported by a PSL post-doctoral fellowship. 
This project has been supported by the European Research Council (ERC) under the European 
Union’s Horizon 2020 research and innovation program (Advanced Grant AstroGeo-885250). 
This work was granted access to the HPC resources of MesoPSL financed by the Region 
Île-de-France and the project Equip@Meso (reference ANR-10-EQPX-29-01) of the 
programme Investissements d’Avenir supervised by the Agence Nationale pour la Recherche. 

\section*{Data Availability}
All data needed to evaluate the conclusions in the paper are present in the paper 
and/or the Appendices.

\bibliographystyle{mnras}
\bibliography{DISS} 


\appendix 

\section{Secular dynamics at second order in planetary masses} 
\label{app:L46}
We use the secular equations of motions of \citep[][and references therein]{laskar1985,laskar1990,laskar2008}. They were obtained via series expansions in planetary masses, eccentricities, and inclinations, as well as through  second-order  analytical averaging over the rapidly changing mean longitudes of the planets. The expansion was truncated at the second order with respect to the masses and to degree 5 in eccentricities and inclinations. The equations include corrections from general relativity and Earth-Moon gravitational interaction. This leads to the following system of ordinary differential equations, denoted by $\mathcal{L}_6$ throughout this paper:
\begin{equation}
\begin{aligned}
\label{eq:SecEq6}
\frac{d \omega}{dt} = \sqrt{-1}  \{ \Gamma +  \Phi_3(\omega, \bar{\omega}) + \Phi_5(\omega, \bar{\omega})  \},
\end{aligned}
\end{equation}
where $\omega = (z_1,\dots,z_8,\zeta_1, \dots, \zeta_8)$, with $z_k = e_k \E^{j \varpi_k}$ and $\zeta_k = \sin(i_k/2) \E^{j \Omega_k} $. 
The planets are indexed in order of increasing semi-major axis, as usual. The variable $\varpi_k$ is the longitude of the perihelion, $\Omega_k$ is the longitude of the ascending node, $e_k$ is eccentricity, and $i_k$ is inclination. The function $\Phi_3(\omega, \bar{\omega})$ and $\Phi_5(\omega, \bar{\omega})$ are the terms of degree 3 and 5, respectively. The $16 \times 16$ matrix $\Gamma$ is the linear Laplace-Lagrange system, which is slightly modified to make up for the higher-order terms in the outer Solar System.

To mimic $\mathcal{H}_4$, we define the new model $\mathcal{L}_4$ by dropping the terms of degree 5 from the equations of the inner planets, 
that is:
\begin{equation}
\label{eq:SecEq4}
\frac{d \omega}{dt} = \sqrt{-1}  \{ \Gamma +  \Phi_3(\omega, \bar{\omega}) + \mathbb{D} \Phi_5(\omega, \bar{\omega})  \}, 
\end{equation}
where we introduced the diagonal matrix $\mathbb{D} = \mathrm{diag}(\vec{0}, \vec{1}, \vec{0}, \vec{1})$, with $\vec{0} = (0,0,0,0)$ 
and $\vec{1} = (1,1,1,1)$. It should be noted that the truncations behind the models 
$\mathcal{L}_6$ and $\mathcal{L}_4$ are defined with respect to the classical variables $z_k$, $\zeta_k$, differently from the models $\Hiss_{2n}$ 
which result from the expansion of $\Hiss$ in the complex Poincaré variables $x_k \propto (1 - (1- e_k^2)^{1/2})^{1/2} \, \E^{j \varpi_k}$ and 
$y_k \propto (1- e_k^2)^{1/4} \sin(i_k/2) \, \E^{j \Omega_k}$. 

We define ensembles of initial conditions by slightly varying a single variable of an inner planet at a time, while keeping other variables identical to their reference values, as shown in Table~\ref{tab:iniCon}. 
For the integrations over 100 Gyr, we use the initial conditions varied from the variables $(k_i = e_i \cos \varpi_i)_{i=1,4} $ of the four inner planets, 
except for the solutions of $\mathcal{L}_4$, where only those varied from $k_1$ are used. For the solutions computed over 5 Gyr, 
the variables $(e_i)_{i=1,4}$ are varied to obtain the initial conditions. 
The solutions integrated up to 100 Gyr are included in the analysis of the statistics of the first 5 Gyr.
Equations~\eqref{eq:SecEq6} and \eqref{eq:SecEq4} are integrated from these ensembles of initial conditions to obtain the solutions of $\mathcal{L}_6$ and $\mathcal{L}_4$. 

\begin{table}
    \centering
    \begin{tabular}{ccccc}
     \hline\hline 
        Variable & Offsets & $\epsilon$ & $N$ & $T$\\
    \hline 
         $k_i$ & $-N\epsilon$ to $ N\epsilon$  &  $10^{-11}$ & 5000 & 100 Gyr\\ 
         $e_i$ &  $-N\epsilon$ to $ N\epsilon$  &  $10^{-11}$ & 10000 & 5 Gyr \\
     \hline
    \end{tabular}
    \caption{Offsets of the initial variables $k_i = e_i \cos \varpi_i$ and eccentricity $e_i$, with $i \in \{1,2,3,4\}$ corresponding to the inner planets $\{$Mercury, Venus, Earth, Mars$\}$. 
    Different initial conditions correspond to offsets of $n \epsilon$ in a single variable of a single planet for $n = - N, \dots , N$, while other variables are kept to their nominal values. 
    Each initial condition is used to compute a solution over the time interval $[0,T]$.}
    \label{tab:iniCon}
\end{table}

\section{Statistics with different thresholds of Mercury's eccentricity} 
\label{app:thresholds}

\begin{figure*}    
\includegraphics[width=\textwidth]{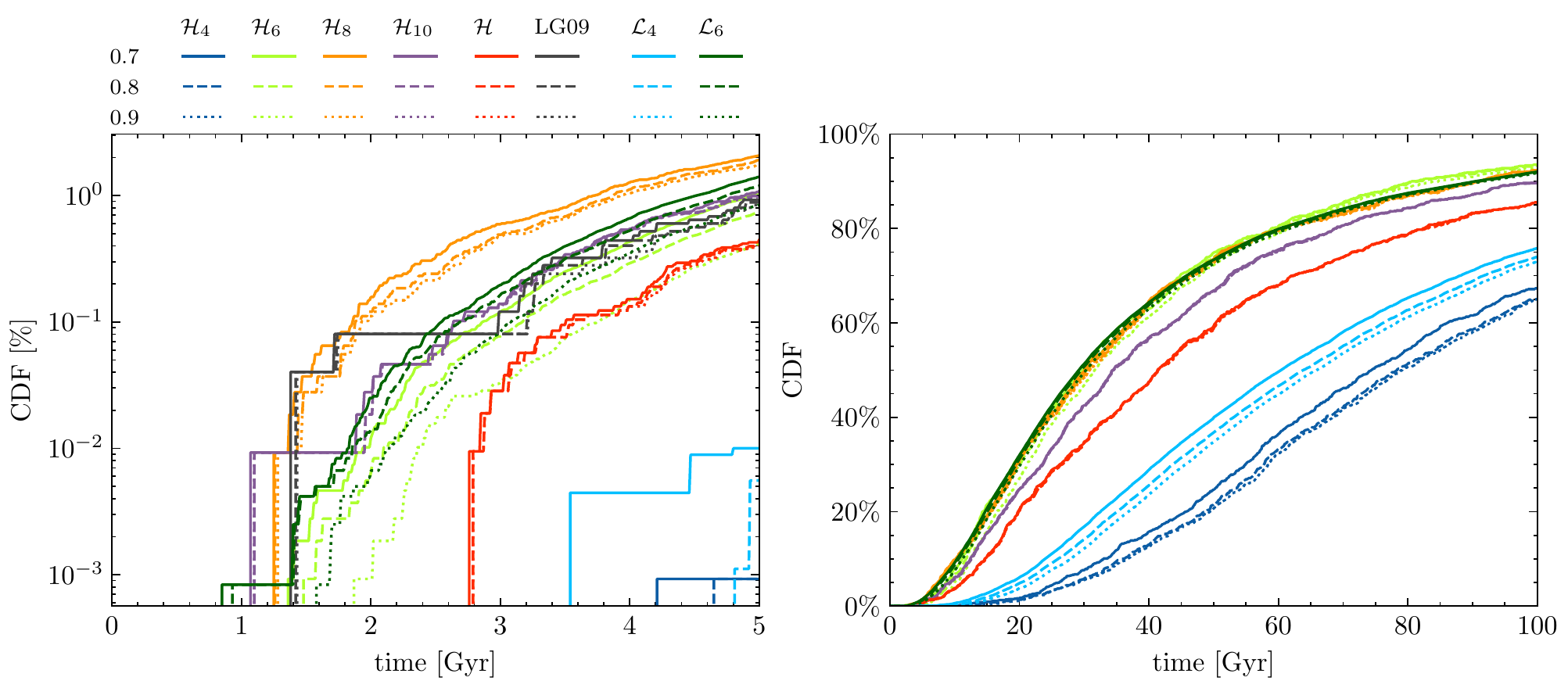}
\caption{CDFs of the first hitting time of Mercury's eccentricity at the three thresholds 0.7, 0.8, and 0.9, up to 5 Gyr (left panel) and 100 Gyr (right panel) 
in the future for the dynamical models $\Hiss_{2n}$, $\Hiss$, $\mathcal{L}_{2n}$, and LG09. LG09 represents 2\,492 orbital solutions over 5 Gyr of \citet{Laskar2009}, 
while $\Hiss$ denotes the 10\,560 and 1\,042 solutions of Gauss' dynamics in \ML{} spanning 5 Gyr and 100 Gyr, respectively. It should be noted that we use 
here the upper bound of estimation for $\Hiss$, in which Mercury's eccentricity of a solution is assumed to exceed 0.9 after a secular collision, as explained 
in the main text.} 
\label{fig:CDFs_compare}
\end{figure*}

We compute the CDFs of the first hitting time of Mercury's eccentricity at the three levels 0.7, 0.8, and 0.9, 
in order to test the dependency of the instability statistics on different thresholds. The results are shown in Fig.~\ref{fig:CDFs_compare}. 
Up to 5 Gyr, when the instability constitutes a rare event, the models of degree higher than 6 show consistency across high values of eccentricity. 
The difference between the CDFs of the three thresholds is relatively significant for the models at degree 6 ($\Hiss_6$, $\mathcal{L}_6$), and even 
more so at degree 4 ($\Hiss_4$, $\mathcal{L}_4$). For $\Hiss_6$ and $\mathcal{L}_6$, only about half of the integrations exceeding 0.7 also goes beyond 0.9 in 5 Gyr. 
It should be noted that if Mercury's eccentricity goes beyond 0.9, it is likely that a catastrophic event will shortly ensue, whether it is a secular 
collision (\ML{}) or a numerical instability in the truncated dynamics. Therefore, the expected time that a solution of $\Hiss_6$ spends in an unstable 
state of high Mercury eccentricity is longer, which makes $\Hiss_6$ a prime model for the study of the unstable states of the ISS. Over a longer timescale 
of 100 Gyr, when the destabilisation is no longer a rare event, the difference of the CDFs with respect to the choice of the eccentricity threshold is small for the models at degree 4 and negligible for the rest. 

\section{Difference between past and future for the statistics of Mercury's eccentricity}
\label{app:PF} 

\begin{figure*}    
\includegraphics[width=\textwidth]{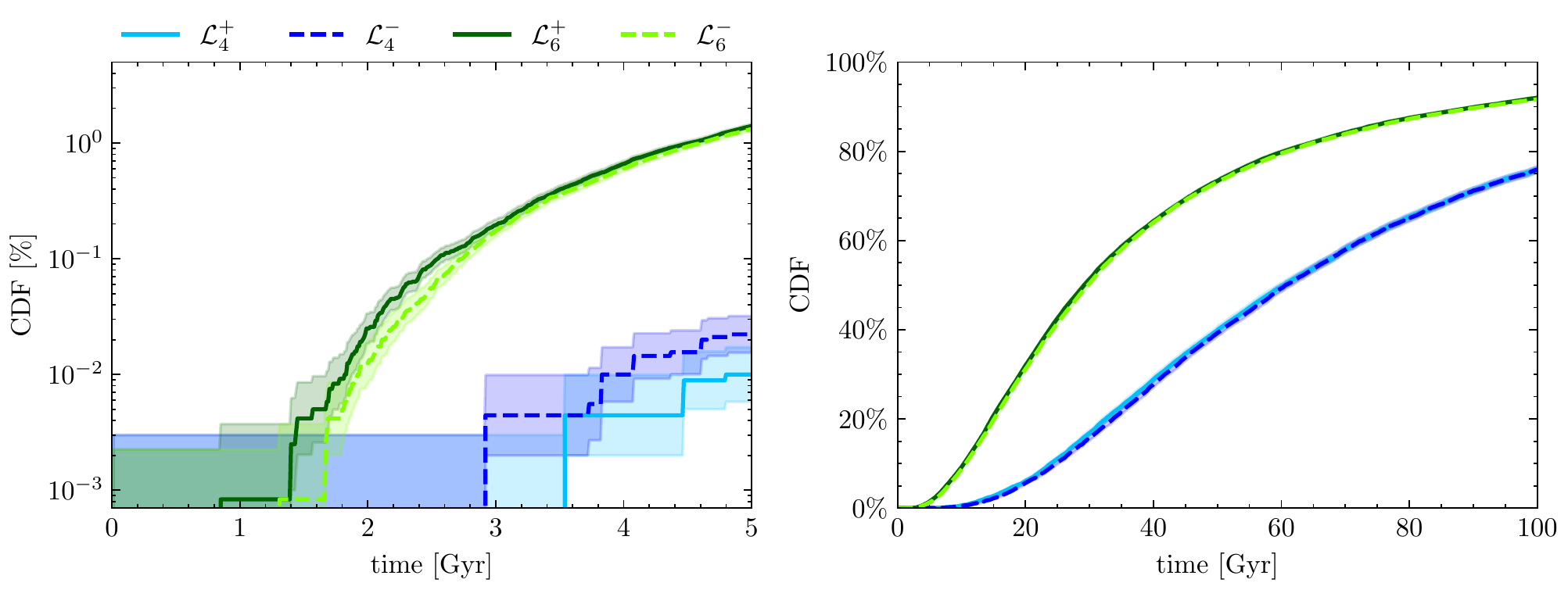}
\caption{CDFs of the first hitting time of Mercury's eccentricity at 0.7 for the dynamical models $\mathcal{L}_{2n}$ over 5 Gyr (left panel) and 100 Gyr (right panel) in the two time directions, 
with 90\% confidence intervals. The integrations in the past are denoted by $\mathcal{L}_{2n}^-$, while $\mathcal{L}_{2n}^+$ denotes the integrations in the future. } 
\label{fig:CDFs_PF}
\end{figure*}

In this work, we focus on the statistics of Mercury's eccentricity over long timescales in the future. It is interesting to revert the time 
direction to obtain the statistics in the past for comparison. From the set of initial conditions defined according to Table~\ref{tab:iniCon}, 
Eqs.~\eqref{eq:SecEq6} and \eqref{eq:SecEq4} are integrated in the direction of negative times to obtain 120\,000 and 40\,000 solutions spanning 
5 Gyr and 100 Gyr, respectively, for both degree 4 and degree 6. 

The CDFs of the first hitting time of Mercury’s eccentricity at 0.7 for $\mathcal{L}_{4}$ and $\mathcal{L}_{6}$ in two time directions are shown in Fig.~\ref{fig:CDFs_PF}. 
For both models, the difference between past and future is small but still noticeable initially, and gradually diminishes as time goes. 
The asymptotic convergence of the two time directions is physically expected, and has been also observed for the PDFs of the fundamental frequencies 
of the Solar System \citep{hoang2021}. Over the timescale of 100 Gyr, the CDFs of past and future are identical for both $\mathcal{L}_{4}$ and $\mathcal{L}_{6}$.

\section{PDF estimation}
\label{supp:PDF}
In this section, we will briefly explain the methods to estimate the PDF of $\tau = \inf_{t} \{ e_1(t) \geq 0.7 \}$, that is, the first time that the eccentricity of Mercury $e_1$ reaches the threshold of 0.7 from our ensembles of integrations spanning from 0 to 100 Gyr. 

\paragraph*{Debiased KDE and bootstrap.}
KDE, also known as the Parzen–Rosenblatt window method, is a non-parametric estimator of the underlying PDF of a dataset \citep{rosenblatt1956,parzen1962}. In this work, we use a bias-corrected version of the KDE to facilitate the uncertainty estimation by bootstrapping the data. We briefly present the method here (a detailed presentation can be found in \citet{cheng2019} and references therein). 
Let $\mathbf{X} = \{ X_1, X_2, \dots, X_n\} $ be a univariate independent and identically distributed (i.i.d.) sample drawn from an unknown probability density function $p(x)$. The KDE of the sample is then defined as:
\begin{equation} 
\label{eq:KDE}
    \widehat{p}_h (x | \mathbf{X} ) = \frac{1}{nh} \sum_{i=1}^n K \left( \frac{x-X_i}{h} \right), 
\end{equation}
where $K$ is a non-negative kernel function and $h$ is the bandwidth. In this work, we choose \citet{silverman1986}’s rule of thumb for the selection of the optimal bandwidth and the standard Gaussian kernel. 
With this choice of bandwidth, the bias error and variance error of the KDE in Eq.~\eqref{eq:KDE} are of the same order of magnitude. Therefore, the bootstrap method \citep{Efron1979}, which measures the variance error by random resampling of the original dataset, is not a consistent estimator of the total error of the KDE in Eq.~\eqref{eq:KDE}. One approach to this problem is to use a bias-corrected KDE, defined as:
\begin{equation}
\label{eq:deKDE}
    \widetilde{p}_h (x) = \widehat{p}_h (x) - \frac{h^2}{2} \sigma^2_{K} \frac{d^2 \widehat{p}_h (x)}{dx^2}, 
\end{equation}
where $\sigma^2_K = \int || x ||^2 K(x) dx$ is a constant depending on the kernel function $K$. With the debiased KDE in Eq.~\eqref{eq:deKDE}, the bias error is reduced so that the total error is dominated by the variance error, which can be consistently estimated by the bootstrap method. 

The procedure of the standard bootstrap \citep{Efron1979} is as follows. We resample the original dataset $\mathbf{X}$ with replacement to obtain a bootstrap sample  $\mathbf{X}^* = \{ X_1^*, X_2^*, \dots, X_n^*\} $. Equation~\eqref{eq:deKDE} is then applied to this bootstrap sample to obtain a bootstrap debiased KDE $\widetilde{p}_h^* (x | \mathbf{X}^* )$. We then repeat this procedure B times to obtain B bootstrap debiased KDEs $\widetilde{p}_h^{*(1)}, \cdots, \widetilde{p}_h^{*(B)} $. Because the distribution of $|\widetilde{p}_h^* - \widetilde{p}_h (x)| $ approximates that of $|\widetilde{p}_h - p(x) | $, from the sample of the B bootstrap KDEs we can compute an asymptotically valid estimation of the piecewise confidence interval $\mathrm{CI}_{1-\alpha}(x)$, defined as:
\begin{equation}
P(|\widetilde{p}_h - p(x) | < \mathrm{CI}_{1-\alpha}(x) ) = 1 - \alpha.
\end{equation}

\paragraph*{Boundary correction.} 
Kernel density estimation of a PDF on a finite interval can be affected by non-negligible bias at the boundaries. 
In our work, the interval is defined by the total integration time, that is, $[0, 100]$~Gyr in Fig.~\ref{fig:100Gyr}. 
The nature of the two boundaries is different, and they should be treated differently.
At $t=0$, the integrations start closely around a nominal value of $e_1 \approx 0.2$, therefore the PDF of the first hitting time of $e_1=0.7$ should be 0 when $t=0$. 
This constraint suggests the log-transformation of the sample before applying the KDE \citep{charpentier2015}. 

The boundary at 100 Gyr has no similar constraints, and we employ a pseudodata method to correct the bias \citep{cowling1996}. 
The idea is to use the original dataset to generate fictitious data outside the interval of interest. 
Let $X_{(1)} < \dots < X_{(n)}$ be the order statistics of the data $X_1, \dots, X_n$ on the interval $[0,1]$. The extra data points generated in the range $ (-\infty, 0)$ are defined by the three-point rule: 
\begin{equation}
\label{eq:3pointrule}
    X_{(-i)} = -6X_{(i)} + 4X_{(2i)} - 3X_{(3i)}. 
\end{equation}
To adapt the upper limit of the interval [0, 100] Gyr to this rule, we simply transform the data as $X_{(i)} \rightarrow (100-X_{(i)})/100$. The pseudodata are then generated according to Eq.~\eqref{eq:3pointrule}, and the ensemble is back-transformed at the end. The number of pseudodata points is taken to be about $10\%$ of the sample size.


\bsp	
\label{lastpage}
\end{document}